\documentclass[prb,amsmath,amssymb,superscriptaddress,twocolumn]{revtex4}
\usepackage{graphicx}
\usepackage{dcolumn}
\usepackage{bm}
\newcommand{\be}{\begin{equation}}
\newcommand{\ee}{\end{equation}}
\newcommand{\ber}{\begin{eqnarray}}
\newcommand{\eer}{\end{eqnarray}}
\newcommand{\He}{$^3$He}
\newcommand{\Hefour}{$^4$He}
\newcommand{\onehalf}{\frac{\mbox{\small 1}}{\mbox{\small 2}}}
\newcommand{\onethird}{\frac{\mbox{\small 1}}{\mbox{\small 3}}}
\newcommand{\onefifth}{\frac{\mbox{\small 1}}{\mbox{\small 5}}}
\newcommand{\twothirds}{\frac{\mbox{\small 2}}{\mbox{\small 3}}}
\newcommand{\tinyonethird}{\frac{\mbox{\tiny 1}}{\mbox{\tiny 3}}}
\newcommand{\grad}{\mbox{\boldmath$\nabla$}}
\newcommand{\taud}{\tau_{\text{\tiny D}}}
\newcommand{\diff}{\text{D}_{\text{\tiny M}}}
\newcommand{\x}{\tilde{x}}
\def\vj{{\bf j}}
\def\vp{{\bf p}}
\def\vM{{\bf M}}
\def\vr{{\bf r}}
\begin{document}
\title{Magnetization and Spin-Diffusion of Liquid \He\ in Aerogel}
\author{J.A.Sauls}
\affiliation{Centre de Recherches sur les Tr\`es Basses Temp\'eratures,
             Centre National de la Recherche Scientifique,\\
             Laboratoire Associ\'e \`a l'Universit\'e J. Fourier,
            BP 166, 38042 Grenoble Cedex 9, France}
\affiliation{Department of Physics \& Astronomy, Northwestern University, Evanston, IL 60208}
\author{Yu. M. Bunkov}
\author{E. Collin}
\author{H. Godfrin}
\affiliation{Centre de Recherches sur les Tr\`es Basses Temp\'eratures,
             Centre National de la Recherche Scientifique,\\
             Laboratoire Associ\'e \`a l'Universit\'e J. Fourier,
            BP 166, 38042 Grenoble Cedex 9, France}
\author{P. Sharma}
\affiliation{Department of Physics \& Astronomy, Northwestern University, Evanston, IL 60208}
\date{\today}
\begin{abstract}
We report theoretical calculations of the normal-state spin diffusion
coefficient of \He\ in aerogel, including both elastic and inelastic scattering
of \He\ quasiparticles, and compare these results with experimental data for
\He\ in 98\% porous silica aerogel. This analysis provides a determination of
the elastic mean free path within the aerogel. Measurements of the magnetization
of the superfluid phase provide a test of the theory of pairbreaking and
magnetic response of low-energy excitations in the ``dirty'' B-phase of \He\ in
aerogel. A consistent interpretation of the data for the spin-diffusion
coefficient, magnetization and superfluid transition temperature is obtained by
including correlation effects in the aerogel density.
\end{abstract}
\maketitle

\section{\label{introduction}Introduction}

A strongly correlated Fermi liquid in the presence of disorder
is realized when \He\ is introduced into
highly porous silica aerogel. The effects of disorder on the phase
diagram and low-temperature properties of superfluid \He\ have been a
subject of widespread current interest. The thermodynamic and transport
properties of \He\ in aerogel are strongly modified by the scattering of
\He\ quasiparticles off the low-density aerogel structure. Here we
summarize theoretical calculations of the magnetization\cite{sha01} and
spin transport\cite{sha03a} in the normal and superfluid states of
\He\ in high-porosity aerogel. The theoretical results are compared with
experimental measurements of the normal-state spin-diffusion coefficient
and magnetic susceptibility of the superfluid phase over the temperature
range $1\,\text{mK}\lesssim T \lesssim 100\,\text{mK}$.\cite{col02}

For $98\%$ porosity the typical diameter of the silica strands is $\delta\simeq
3\,\text{nm}$. The mean distance between strands is $\xi_a\simeq 40\,\text{nm}$,
which is large compared to the Fermi wavelength, $\lambda_f\sim 0.1\,\text{nm}$,
but comparable to or less than the bulk coherence length, $\xi_0=\hbar v_f/2\pi
T_{c0}\simeq 20-80\,\text{nm}$ over the pressure range $p=34-0\,\mbox{bar}$. The
aerogel does not modify the bulk properties of normal $^3$He, beyond the
formation of a couple of atomic layers of solid-like $^3$He adsorbed on the
silica strands. The main effect of the aerogel structure is to scatter \He\
quasiparticles moving with the Fermi velocity. At temperatures below $T^*\approx
10\,\text{mK}$ elastic scattering by the aerogel dominates inelastic
quasiparticle-quasiparticle collisions.\cite{rai98} This limits the mean free
path of normal \He\ quasiparticles to $\ell\simeq 130-180\,\text{nm}$ for
aerogels with $98\%$ porosity. (Note that there are variations in mean free path
and correlation length for different aerogels with the same porosity.) Thus, the
low-temperature limit for the spin diffusion coefficient is determined by elastic
scattering from the aerogel. Comparison with experimental measurements for $T\ll
T^*$ provides a determination of the transport mean free path due to scattering
of quasiparticles by the aerogel.

\begin{figure}
\includegraphics[width=4cm]{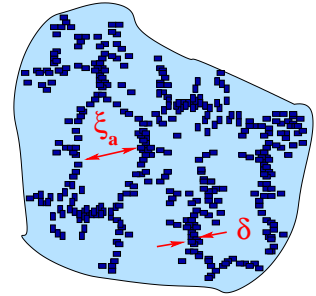}
\caption{\label{fig:aerogel_model} A model silica aerogel showing
low-density regions (light) of \He\ surrounded by higher density strands
and aggregates of silica (dark). The aerogel correlation length,
$\xi_a\approx 30$ nm, is identified with the mean inter-strand distance.
The aerogel strand diameter is approximately, $\delta\simeq 3$ nm.}
\end{figure}

\begin{figure}
\includegraphics[width=6cm]{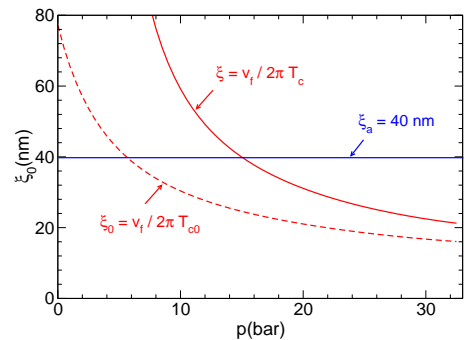}
\caption{\label{fig:length-scales} The pair correlation length of
superfluid \He\ in aerogel (solid curve) as a function of pressure is shown in comparison
with an aerogel strand-strand correlation length, $\xi_a\simeq 40\,\text{nm}$.
A cross-over occurs near $p\approx 15\,\text{bar}$. The bulk \He\ correlation length
is also shown (dashed curve).}
\end{figure}

Scattering by the aerogel also has dramatic effects on the
formation and properties of the superfluid phases. If the coherence length
(pair size) is sufficiently long compared to the typical distance between
scattering centers, then the aerogel is well described by a homogeneous,
isotropic scattering medium (HSM) with a mean-free path determined by the
aerogel density. However, more elaborate scattering models are required
if aerogel density correlations develop on length scales that are
comparable with the pair correlation length.\cite{thu98} Density
correlations are observed at wavevectors $q\gtrsim \pi/\xi_a$ in the
aerogel structure factor. We identify $\xi_a$ with the typical distance
between silica strands or clusters, $\xi_a\approx 30-50\,\text{nm}$
(c.f. Fig. \ref{fig:aerogel_model}).
Figure \ref{fig:length-scales} shows a comparison of the
pair correlation length (Cooper pair size), $\xi = \hbar v_f/2\pi T_c$,
as a function of pressure for \He\ in aerogel, as well as that for pure \He\
($\xi_0 = \hbar v_f/2\pi T_{c0}$), with an aerogel correlation length of
$\xi_a=40\,\text{nm}$.

The HSM provides a reasonable approximation to the properties of
superfluid $^3$He-aerogel at low pressures; this model accounts
semi-quantitatively for the reduction of $T_c$, including the
critical pressure, $p_c$, and the pair-breaking suppression of the order
parameter.\cite{thu98,rai98} However, the HSM becomes a poorer
description of $^3$He-aerogel at higher porosities and higher pressures where
the pair size is comparable to, or smaller than, the typical distance between
scattering centers. This breakdown of the HSM is first evident in the
quantitative discrepancies in the pressure dependence of $T_c$ particularly
for higher porosity aerogels.\cite{thu98,law00}

The inhomogeneity of the aerogel on scales $\xi_a\gtrsim\xi$ leads to
higher superfluid transition temperatures than predicted by the HSM with
the same quasiparticle mean free path. Regions of lower aerogel density, of
size of order $\xi_a$, are available for formation of the condensate. Thus,
the qualitative picture is that of a random distribution of low density
regions, `voids', with a typical length scale, $\xi_a$, in an aerogel with
a quasiparticle mean-free path, $\ell$. In the limit $\xi\sim\xi_a\ll\ell$,
the superfluid transition is determined by the pairbreaking effects of
dense regions surrounding the `voids', and scales as $\delta
T_c/T_{c0}\propto -(\xi/\xi_a)^2$. However, when the pair size is much
larger than $\xi_a$ the aerogel is effectively homogeneous on the scale of
the pairs and pairbreaking results from homogeneous scattering defined by
the transport mean free path, which scales as $\delta T_c/T_{c0}\propto
-(\xi/\ell)$. This latter limit is achieved at low pressures. Here we adopt
a simplified version of the \textsl{inhomogeneous} scattering model\cite{thu98}
that incorporates correlations of the aerogel. The correlation
effect is included by introducing an effective pairbreaking parameter that
interpolates between these two limits: $x\to\x=x/(1+\zeta_a^2/x)$, where
$x=\xi/\ell$ and $\zeta_a\equiv\xi_a/\ell$. This heuristic treatment of
aerogel correlations provides a good description of the pressure dependence
of $T_c$ in zero field for \He\ in aerogel over the whole pressure
range.\cite{sau03}

In section \ref{Magnetic_susceptibility} we summarize experimental
results for the magnetization of normal \He\ in aerogel which are used
to determine the amount and effect of the solid-like component of \He\
that coats the aerogel strands. Section \ref{spin_diffusion} describes
NMR measurements of the spin diffusion coefficient of normal liquid \He\
in aerogel and the theoretical calculations used to determine the
elastic mean free path of quasiparticles. In p-wave superfluids
quasiparticle scattering from the aerogel medium is intrinsically
pairbreaking and leads to renormalization of nearly all properties of
the superfluid phases. In section \ref{magnetic_susceptibility} we
summarize an analysis of the effects of scattering by the aerogel on the
magnetic susceptibility based on the theoretical results for the
disordered Balian-Werthamer (BW) phase,\cite{sha01} modified to include
the reduced pair-breaking effect of low-density regions of aerogel. The
``dirty'' B-phase is believed to describe the zero-field phase of
superfluid \He\ in aerogel, except in a narrow region near
$T_c$.\cite{ger02c} The theory for the magnetization based on the
``dirty'' B-phase is tested against the measurements of the
susceptibility.\cite{col02}
\section{\label{Magnetic_susceptibility}
                Magnetization of normal $^3$He}

The nuclear magnetization, $M$, of normal liquid $^3$He at
temperatures, $k_B T\ll E_f$, and fields, $\gamma\hbar H \ll E_f$,
is given in terms of the (single-spin) quasiparticle density of
states at the Fermi level, $N_f$, the nuclear gyromagnetic ratio for
$^3$He, $\gamma$, and the exchange enhancement of the local field
given in terms of the Landau interaction parameter, $F_0^a$, $
\chi_N = M/H  = 2N_f\mu^2/(1+F_0^a)$, where
$\mu=\gamma\hbar/2$ is the nuclear magnetic moment of the $^3$He
nucleus; $\chi_N$ is the nuclear spin susceptibility of the normal
Fermi liquid.

The effect of the aerogel on the magnetization of the normal {\sl
liquid} phase of $^3$He is expected to negligible. However, the
aerogel structure is known to be covered with one or two layers of
localized $^3$He atoms. These surface layers contribute a Curie-like
susceptibility that obscures the Fermi-liquid contribution at low
temperatures.\cite{spr95} The surface contribution can be
suppressed by the addition of $^4$He which
preferentially plates the aerogel structure. The net effect is
two-fold: (1) the surface Curie susceptibility is suppressed and
(2) spin-spin scattering between $^3$He quasiparticles and the
surface spins is suppressed. The cross-section of the aerogel may
also be modified by $^4$He pre-plating, but we expect this effect to
be relatively small.

\begin{figure}
\includegraphics[width=8cm]{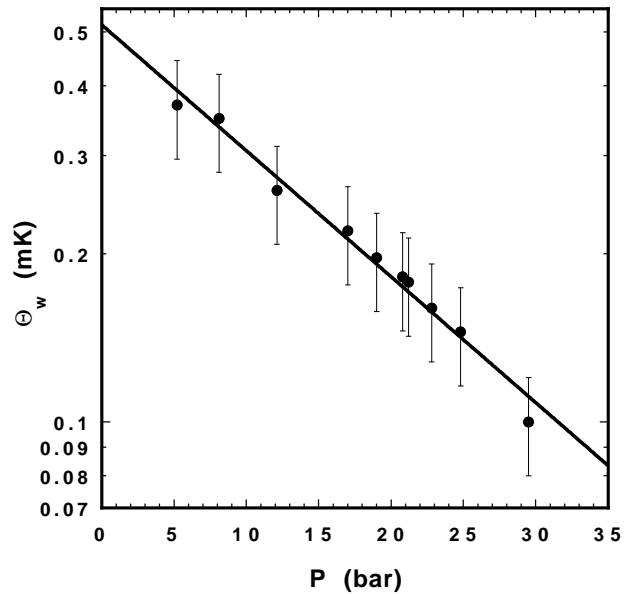}
\caption{\label{fig:Theta} The Curie-Weiss temperature, $\Theta_{\text{\tiny W}}$,
as function of pressure. The decrease reflects the increased average density
of the solid layer in equilibrium with liquid \He\ as the pressure increases.}
\end{figure}

We have performed accurate measurements of the magnetic susceptibility
of normal liquid \He\ in aerogel, with and without $^4$He pre-plating, using
CW NMR at a field of $H=37.3$ mT over a wide range of pressures and
temperatures from $T\simeq 1$ mK up to a few hundred mK. The
temperature was measured using the vibrating wire technique described
in Ref. \onlinecite{win04}. The high temperature regime was calibrated against the
Fermi temperature, $T_F$, and the melting curve
thermometer.
\begin{figure}
\includegraphics[width=8.5cm]{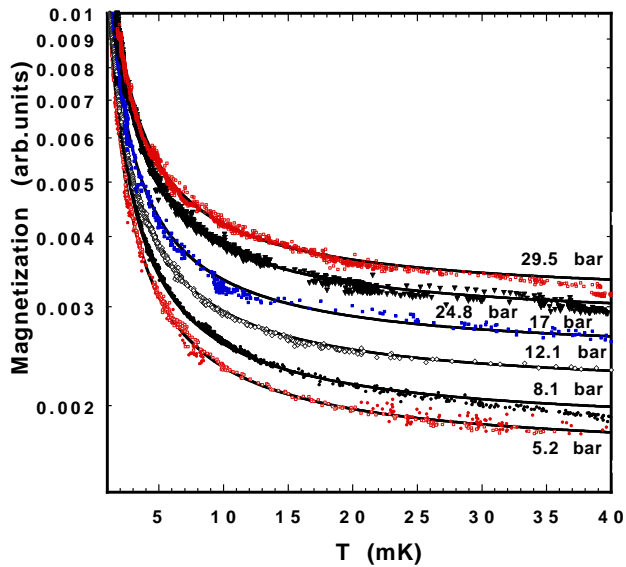}
\caption{\label{fig:Magnetiz}
The total magnetization of \He\ in aerogel, both liquid and solid,
for several pressures.}
\end{figure}
In Fig. \ref{fig:Magnetiz} we show the magnetization data as a function
of temperature for several pressures. The data is fit to a two-component
form for the susceptibility; a Curie-Weiss term to describe the solid
susceptibility with ferromagnetic correlations
plus a temperature-independent term for the bulk Fermi
liquid, $\chi=C/(T-\Theta_{\text{\tiny W}}) + \chi_{\text{N}}$. The fit
allows us to determine the ratio of the number of solid to liquid \He\
atoms as a function of pressure, as well as the Curie-Weiss temperature
of the solid spins.
\begin{figure}
\includegraphics[width=8cm]{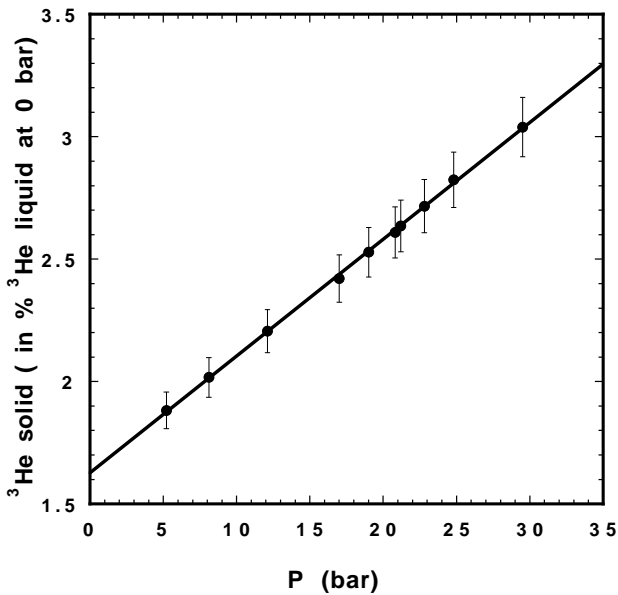}
\caption{\label{fig:solid} The density of solid \He\ atoms,
         normalized by density of liquid \He\ for $p=0$ bar.}
\end{figure}
Pressurizing \He\ in aerogel leads to increased
solidification of \He\ atoms, as shown in Fig. \ref{fig:solid} and a
decrease in the Curie-Weiss temperature as shown in Fig.
\ref{fig:Theta}. Both effects are well understood from previous studies on
\He\ adsorbed on homogeneous substrates (e.g. graphite) or heterogeneous
substrates (powders, Vycor, etc.).\cite{god95,gol96}
These results allow us to remove the contribution to
the susceptibility from solid $^3$He, and thus, to extract the
susceptibility of liquid $^3$He below superfluid transition.

We also performed measurements of the susceptibility of \He\ in aerogel
by pre-plating with $^4$He atoms. We added a small amount of $^4$He at a
temperature of about $3$ K and then cooled down. The solid \He\
susceptibility is used to measure the amount of $^4$He adsorbed on the
aerogel strands. The Curie-Weiss temperature is found to decrease in
proportion to the amount of residual solid $^3$He.

For the NMR experiments reported here the solid and liquid components of
$^3$He in aerogel exhibit a
common NMR line, and are in the limit of ``fast exchange'' with a resonance
frequency determined by the weighted average of
the liquid and solid resonance frequencies,
$\langle\omega\rangle=
\left(M_{\text{\tiny liquid}}\,\omega_{\text{\tiny liquid}}+
M_{\text{\tiny solid}}\,\omega_{\text{\tiny solid}}
\right)/\left(M_{\text{\tiny liquid}}+M_{\text{\tiny solid}}\right)$.\cite{spr95}
We have performed the measurement of first and second moments of the NMR
line, with and without $^4$He pre-plating. The NMR line is broadened by the
presence of solid \He\ by as much as 12 $\mu$Tesla, and the broadening
scales by the ratio of the solid to liquid magnetization.

These experiments combined with NMR measurements of the spin-diffusion process in
a field gradient provide a detailed characterization of the magnetic properties
of solid and liquid $^3$He in the aerogel sample. The determination of the
solid \He\ magnetization, and the extrapolation of the Curie-Weiss behavior
to temperatures below the superfluid transition allow us to extract the liquid
component of the susceptibility in the superfluid phase of \He\ in aerogel.
Measurements of the spin-diffusion coefficient in the normal phase of \He\ provide
a measurement of the transport mean free path in aerogel.

\section{\label{spin_diffusion}Spin Diffusion}

Measurements of the spin-diffusion coefficient for \He\ in $98\%$
aerogel were performed with small amounts of $^4$He added in
order to displace the solid \He. At $p=0$ bar we see no signature
of solid $^3$He, while at $p=29.5$ bar we observe a small fraction
($\approx 0.25$\% in the units of Fig. \ref{fig:solid}) of solid \He.

\begin{figure}
\includegraphics[width=8cm]{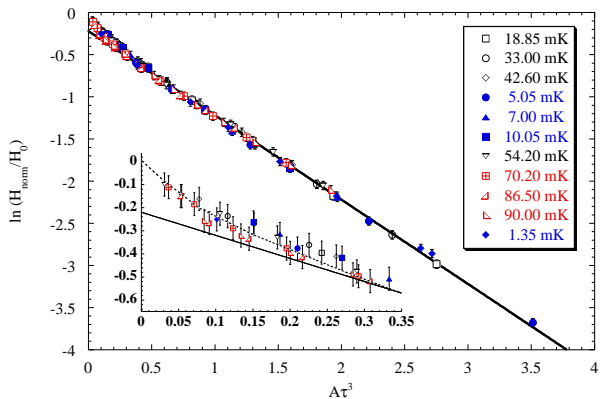}
\caption{\label{fig:tkub}
The $\tau^3$ decay of the echo signal in a magnetic field gradient for $p=0.5$
bar. The temperature range is from $T=1$ to $90$ mK. The coefficient, $A$,
determined from the fit at a fixed temperature is related to the
spin-diffusion coefficient and field gradient. The inset shows the short
time, fast relaxation of unknown origin.}
\end{figure}

We have used the typical spin echo sequences, $\pi/2 - \tau -\pi$ and
$\pi/2 - \tau -\pi/2$,
to measure the decay of echo signal as a function of
time delay $\tau$ at a magnetic field gradient $G_z =
23\mu$T/cm. The solution of the Bloch equations for
the spin echo signal in the magnetic field gradient reads is given by
$H=H_\circ\exp{(-2\tau/T_2-A\tau^3)}$, where $A = \twothirds \diff(\gamma G_z)^2$.
The decay of the spin echo is shown in Fig. \ref{fig:tkub} as function of
$A\tau^3$ for our experimental conditions. On short time scales we observe
small deviations from the $\tau^3$ dependence. These
deviations are indicative of a fast relaxation mechanism of unknown
origin. However, the main feature of the data is the $\tau^3$ dependence due to
spin diffusion in a magnetic field gradient. The echo decay thus provides a
direct measure of the spin diffusion coefficient
of \He\ in aerogel.

In pure \He, the transport of magnetization in the hydrodynamic limit
is given by the magnetization current density,
\begin{equation}
\vj_i = -\diff\,\grad\,M_i
\end{equation}
where $M_i$ is the $i^{\text{\tiny th}}$ component of the
local magnetization, $\vM(\vr,t)$, and $\diff$ is the spin
diffusion coefficient. The hydrodynamic coefficient can be
calculated from a kinetic theory of quasiparticles, which in the
case of \He\ in aerogel is the Landau-Boltzmann transport
equation with the collision integral determined by both elastic
scattering from the aerogel medium and inelastic quasiparticle
collisions, i.e. binary collisions of quasiparticles within the
narrow energy shell of order $k_B T$ near the Fermi surface.
The general solution has the form,
\be\label{DM}
\diff=\onethird v_f^2(1 + F_0^a)\taud
\,,
\ee
in the hydrodynamic limit, $\omega_{\text{\tiny L}}\ll\taud^{-1}$,
where $\omega_{\text{\tiny L}}=\gamma H$ is the Larmor frequency,
$\taud^{-1}$ is the collision rate that limits the transport of
magnetization, $v_f$ is the Fermi velocity and $F_0^a$ is the $l=0$
exchange interaction for \He\ quasiparticles. The collision rate is
calculated from the solution of the Landau-Boltzmann transport
equation including both scattering channels.\cite{sha00} The
result of this calculation is summarized below.

At high temperatures (${T\gg T^*}$) inelastic binary collisions of
quasiparticle limit the transport of spin. The diffusion time approaches
the bulk transport time,\cite{bro68}
\be
\label{tauD_bulk}
\taud\rightarrow\taud^{\text{\tiny bulk}}=
\frac {2\tau_{\text{\tiny in}}}{\pi^2}\,
\sum_{\nu = 1}^{\text{\tiny odd}}\,
\frac{(2\nu + 1)}{\nu(\nu+1)[\nu(\nu+1) - 2\lambda_D]}
\,,
\ee
where the inelastic quasiparticle lifetime is given by
\be
\label{tau_in}
\tau_{\text{\tiny in}}^{-1}=\frac{N_f^2}{2p_fv_f}\langle{W}\rangle\,T^2
\,,
\ee
with $N_f$ the single-spin density of states at the Fermi energy,
and $W$ the binary collision probability for quasiparticles on the
Fermi surface with momenta, $|\vp_i|=p_f$. The Fermi-surface average
is defined by
\be
\langle{W}\rangle=\int{\frac{d\Omega}{4\pi}}\,
\frac{W(\theta,\phi)}{2\cos(\theta/2)}
\ee
where $W(\theta,\phi)$ is the scattering probability for binary collisions
of quasiparticles on the Fermi surface, defined in terms of the standard
scattering angles.\cite{baym91} The parameter $\lambda_D$
is given by the following average of the spin-flip scattering rate,
\be
\lambda_D=1-\frac{1}{\langle{W}\rangle}\,
\int\frac{d\Omega}{4\pi}\,
\frac{W_{\uparrow\downarrow}\,(1-\cos\theta)\,(1-\cos\phi)}{4\cos(\theta/2)}
\,.
\ee

However, at sufficiently low temperatures the
scattering rate is limited by the quasiparticle collisions with the
aerogel medium. Hence, as the temperature is reduced, the diffusion
coefficient crosses over from a clean-limit behavior, $\diff\propto
T^{-2}$, to an impurity-dominated regime in which the diffusion
coefficient approaches a temperature-independent value given by
\be
\label{tauD_aero}
\taud\rightarrow\tau_{\text{\tiny el}}=\ell/v_f\,,\quad{T\ll T^*}
\,,
\ee
where $\ell$ is the limiting mean-free path for quasiparticles
propagating ballistically through \He\ then scattering elastically
off the aerogel structure. This is the geometric mean free path for
classical (point) particles, i.e. quasiparticles, travelling through
the aerogel medium.

At intermediate temperatures the temperature dependence of the
diffusion time is determined by both scattering channels. The solution
to the Landau-Boltzmann transport equation\cite{sha00} gives,
\ber
\label{tauD}
\taud=\frac{\tau_{\text{\tiny in}}}{4\pi^2}\,\sum_{N=0}^{\text{\tiny even}}\,
\frac{1}{1-\lambda_D/\alpha_N}\times\,
\nonumber
\\
\left\{\sum_{l=0,2,...}^{N}\,
       d_l\,\beta(\gamma+\onehalf;\frac{l+1}{2})\right\}^2
\,,
\eer
where
$\alpha_N=(\gamma+N)\,(1+2\gamma)+\frac{N(N-1)}{2}$, the Beta
function, $\beta(\mu;\nu)=\Gamma(\mu)\Gamma(\nu)/\Gamma(\mu+\nu)$, is related to
the standard Gamma function,\cite{abramowitz70} and
\be
\gamma=\onehalf
\sqrt{1+\frac{1}{\pi^2}\frac{\tau_{\text{\tiny in}}}{\tau_{\text{\tiny el}}}}
\,,
\ee
is the dimensionless parameter that controls the cross-over from
elastic- to inelastic-dominated scattering. The coefficients in
Eq.({\ref{tauD}}) are obtained from the recursion relation,
\be
d_{l+2}=\frac{l(l-1)+2l(1+2\gamma)-2\alpha_N+2\gamma+4\gamma^2}{(l+2)(l+1)}\,d_l
\,,
\ee
and the normalization condition
\be\label{norm}
\hspace*{-3mm}\begin{array}{ll}
\displaystyle{1}
&
\displaystyle{=\sum_{l=0,2,...}^N\sum_{l'=0,2,...}^N \,d_l\,d_{l'}\,,\times}
\\
&
\Big\{
2\gamma^2\,\beta(2\gamma;\frac{l+l'+1}{2})
+
2\gamma^2\,\beta(2\gamma;\frac{l+l'+3}{2})-
\\
&
2\gamma l'\,\beta(2\gamma+1;\frac{l+l'+1}{2})
+
\onehalf\,ll'\,\beta(2\gamma+2;\frac{l+l'-1}{2})
\Big\}
\,.
\end{array}
\ee

\begin{figure}
\includegraphics[width=8cm]{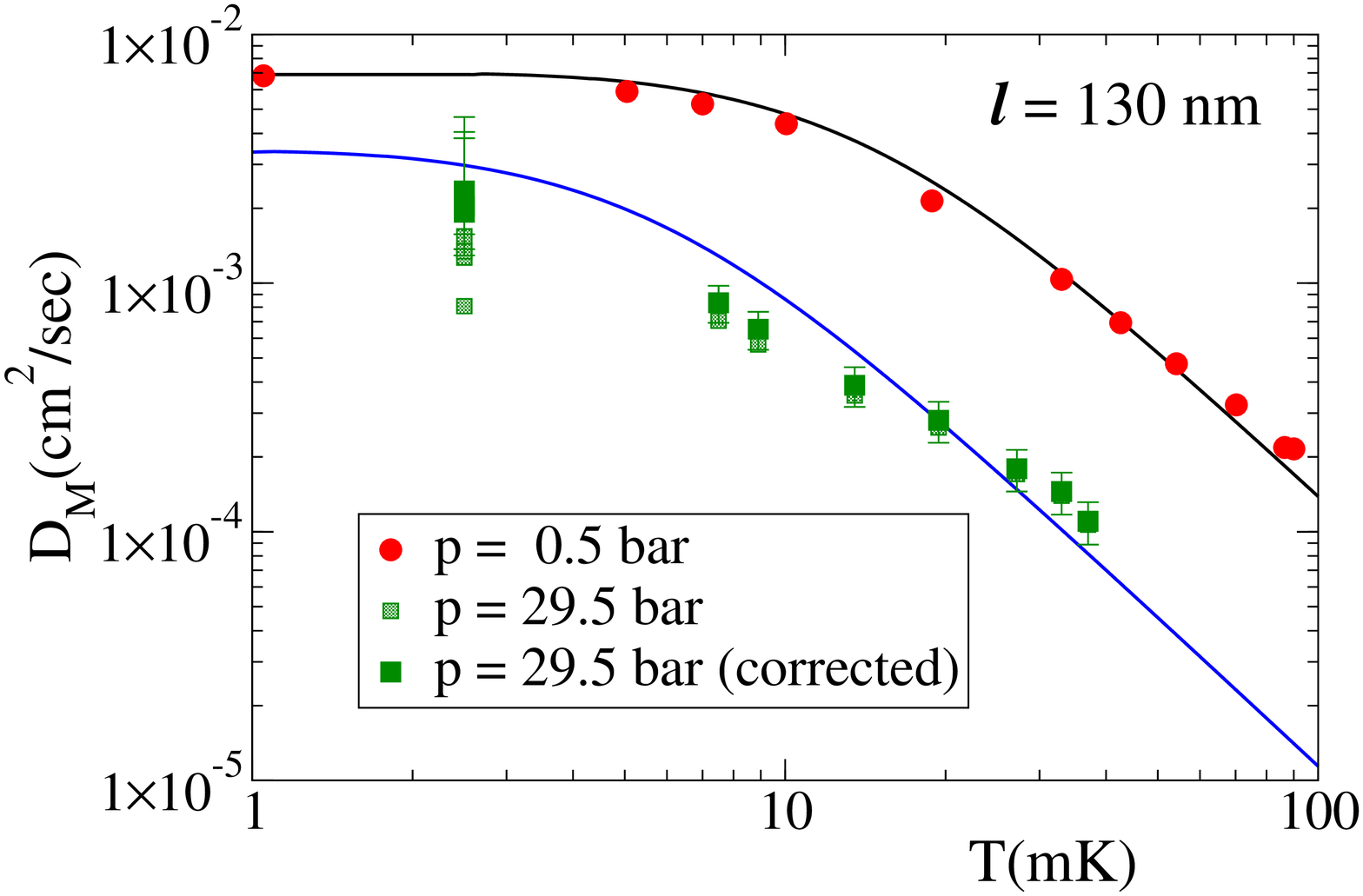}
\caption{\label{fig:GrenobleCompare} Comparison to our experimental data.\cite{col02}
We fit the inelastic scattering amplitude to the
high-temperature data and obtain good agreement for a mean free path $\ell=130$ nm
for the entire temperature range.}
\end{figure}

The calculated spin diffusion coefficient is shown in
Fig. \ref{fig:GrenobleCompare} for $p=0.5\,\text{bar}$ and for
$p=29.5\,\text{bar}$. The spin-diffusion coefficient
decreases as $\diff\propto T^{-2}$ at high temperatures and coincides
with the bulk measurements of the \He\ spin-diffusion coefficient by Ref.
\onlinecite{and62}. Thus, we fit the average binary quasiparticle collision
probability, $\langle{W}\rangle$, at each pressure to the
high-temperature diffusion coefficient with the spin-flip
scattering rate parameter, $\lambda_{\text{\tiny D}}$, obtained from
the scattering model of Ref. \onlinecite{sau81b}. The bulk Fermi liquid parameters
used in the calculation of $\diff$ for \He\ in aerogel are taken from
Ref. \onlinecite{hal90}.

The low-pressure results provide the best data set to compare with the
theory. The theoretical result shown in Fig. \ref{fig:GrenobleCompare}
corresponds to a low-temperature (elastic) mean free path of
$\ell=130\,\text{nm}$. Note that the cross-over from the high-temperature
inelastic limit to the low-temperature aerogel-dominated regime is well
described by the theory at low pressure.

The aerogel mean-free path is
expected to be essentially pressure independent, so long as the integrity
of the aerogel is not damaged under hydrostatic pressure. Thus, the
calculation of the spin-diffusion coefficient is made with the same aerogel
mean-free path and the bulk scattering parameters appropriate for
$p=29.5\,\text{bar}$. The result is also shown in Fig.
\ref{fig:GrenobleCompare}. The experimental data are consistent with the
theoretical results within the estimated errors. The higher pressure spin-diffusion
data is also shown corrected for the small, but measurable, component of
solid magnetization that is transported by diffusion due fast exchange
with the liquid. The main result of this
analysis is the determination of the elastic mean free path. This parameter
is relevant to the mechanism for pairbreaking, the low-energy excitation
spectrum and polarizability of the superfluid phase of \He\ in aerogel.

\section{\label{magnetic_susceptibility}Magnetization of \He-B-aerogel}

Measurements of the B-like suppression of the susceptibility in
$^3$He-aerogel have been reported for two or more monolayers of $^4$He added
to displace the solid layer of $^3$He.\cite{spr95,bar00} Experiments carried
out in Grenoble\cite{col02} on pure \He\ in aerogel also show the B-like
suppression of the liquid component of the susceptibility, obtained by
subtracting the Curie-Weiss component from the solid \He\ coating the aerogel
strands. In the following we compare theoretical results for the magnetic
susceptibility of the ``dirty B phase'' of superfluid \He\ with these recent
experimental measurements. The magnetization measurements in the superfluid
phase were carried out on the same aerogel used in measurements of the spin
echo decay. Thus, the transport mean free path of this aerogel, $\ell =130$
nm, is known.

In the theoretical analysis of the magnetization we consider only elastic
scattering of $^3$He quasiparticles off the aerogel structure, and we neglect the
spin-dependence of the scattering cross-section for \He\ quasiparticles with by
the polarized solid. The splitting of the cross-section by the polarized solid is
expected to be a very small effect which is unimportant in calculating
pair-breaking effects, except in a very narrow temperature
interval corresponding to the $A_1$-$A_2$ transition.\cite{sau03}

The nuclear spin susceptibility of pure superfluid $^3$He-B agrees
quantitatively to leading order in $T_c/E_f$ with the result of
Ref. \onlinecite{ser83} for the susceptibility of the Balian-Werthamer state,
{\small
\be
\label{chi-bulk}
\chi_B/\chi_N = \frac{(1+F_0^a)\left[\twothirds + Y(\onethird + \onefifth F^a_2)\right]}
                     {1+F_0^a(\twothirds+\onethird Y)
             +\onefifth F^a_2(\onethird+\twothirds Y)
                     +\onefifth F^a_2 F^a_0 Y}
\,,
\ee
}
\noindent where $Y(T)$ is the well-known Yosida function,
\be
\label{Yosida}
Y(T) = 1 - \pi T \sum_{\varepsilon_n}\,
\frac{\Delta^2}{\left[\varepsilon_n^2+\Delta^2\right]^{3/2}}
\,,
\ee
$\Delta(T)$ is the B-phase gap amplitude, and $F_2^a$ is the $l=2$ harmonic of
the exchange interaction.

For a B-like phase of \He\ in aerogel depairing of the $S_z=0$ Cooper pairs by
scattering from the aerogel leads to sizeable changes in the magnetization. In
addition to the suppression of $T_{c}$ relative to $T_{c0}$, the magnitude of the
susceptibility, particularly at low temperatures, is sensitive to the density of
quasiparticle states below the gap, $\varepsilon<\Delta$, produced by pairbreaking.

In the HSM model, for either Born
or unitarity scattering, the generic form of Eq. (\ref{chi-bulk}) for the B-phase
susceptibility is preserved with the replacement of the gap and Yosida functions
by impurity-renormalized gap and response functions.\cite{sha01,min02}
The results can be summarized by Eq. (\ref{chi-bulk}) with the replacement of
$Y(T)\rightarrow \tilde{Y}(T)$. For example, in the unitary scattering limit,
{\small
\ber
\label{Y-unitarity}
\hspace*{-5mm}
\tilde{Y} &=& 1 - \pi T\sum_{n}\,
                     \frac{\Delta^2}{\left[\tilde{\varepsilon}_n^2+\Delta^2\right]^{3/2}}
         \Bigg\{\frac{1}{1+ \tinyonethird
                       \left(\frac{1}{\tilde{\varepsilon}_n}\right)^2
               \frac{\Gamma_N\Delta^2}{\sqrt{\tilde{\varepsilon}_n^2+\Delta^2}}}
    \Bigg\}
,
\eer
}

\noindent where $\Gamma_N$ is related to the aerogel mean-free path
for normal-state $^3$He quasiparticles
\be
\Gamma_N=\frac{\hbar v_f}{2 \ell}
\,.
\ee
The gap equation and renormalized Matsubara frequencies are
given by
\be
\ln(T/T_{c0}) = \pi T\sum_{n}\,\left(\frac{1}{\sqrt{\tilde\varepsilon_n^2+\Delta^2}} -
\frac{1}{\vert\varepsilon_n\vert}\right)
\,,
\ee
\be
\tilde\varepsilon_n = \varepsilon_n +
\Gamma_N
\,\tilde\varepsilon_n\sqrt{\tilde\varepsilon_n^2+\Delta^2}/
\tilde\varepsilon_n
\,,
\ee
with $\varepsilon_n = (2n+1)\pi T$.
In the HSM model the gap and excitation spectrum are determined by the
mean free path, $\ell$.
Pairbreaking occurs homogeneously throughout the liquid. The dimensionless pairbreaking
parameter that determines $T_c$ is given by the ratio, $x_c = \Gamma_N/2\pi T_c$. At lower
temperatures the pairbreaking parameter that enters the gap equation and renormalized
Matsubara frequencies scales as $x = \Gamma_N/2\pi T$. Detailed discussion of the results
for the magnetization within the HSM are given in Refs. \onlinecite{sha01,min02}.

When the inhomogeneities of the aerogel occur on length scales comparable to or
greater than the pair correlation length the HSM model overestimates the magnitude
of pairbreaking. For a fixed mean free path this aerogel correlation effect leads
to an increase in $T_c$, and to a reduction in the number of low-energy excitations
produced by scattering within the aerogel. We introduce the effective pairbreaking
parameter for the gap and excitation spectrum based on a similar scaling procedure
introduced for the transition temperature.\cite{sau03} For $T_c$ the effective
pairbreaking parameter, $\tilde{x}_c=x_c/(1+\zeta_a^2/x_c)$, interpolates between
the HSM model with $\tilde{x}_c\propto -(\xi/\ell)$ valid for $\xi\gg\xi_a$ and
the inhomogeneous limit with $\tilde{x}_c\propto -(\xi/\xi_a)^2$ valid for
$\xi\ll\xi_a$. At lower temperatures the scaling is given by
$\tilde{x}=x/(1+\zeta_a^2/x_c)$, which coincides with  $\tilde{x}_c$ for $T\to
T_c$, and accounts for the corresponding reduction in the pairbreaking density of
states at temperatures, $T\ll T_c$. Replacing the pairbreaking parameter of the HSM
by the effective pairbreaking parameter leads to a modification of the low-energy
excitation spectrum and magnetization. Physical properties that are not spatially
resolved on the scale of $\xi_a$ should be well described by
this approach.

\begin{figure}
\includegraphics[width=8cm]{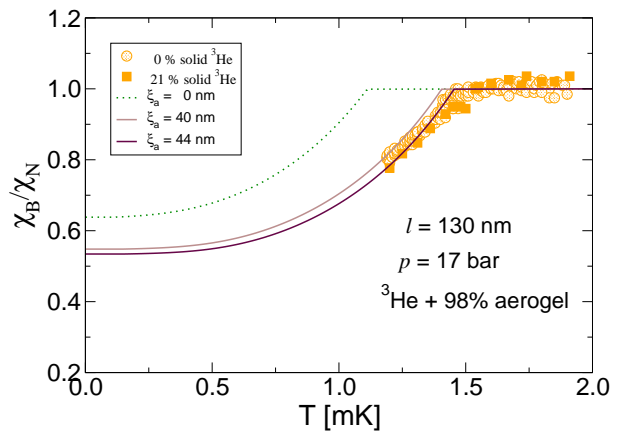}
\caption{\label{fig:GrenobleCompareChi17bar} Comparison to
experimental data for the magnetic susceptibility at
$p=17\,\text{bar}$. Only the liquid contribution to the
susceptibility is plotted for the two measurements with different
amounts of solid \He. The theoretical calculation is based on the
HSM and the dirty BW phase.}
\end{figure}

\begin{figure}
\includegraphics[width=8cm]{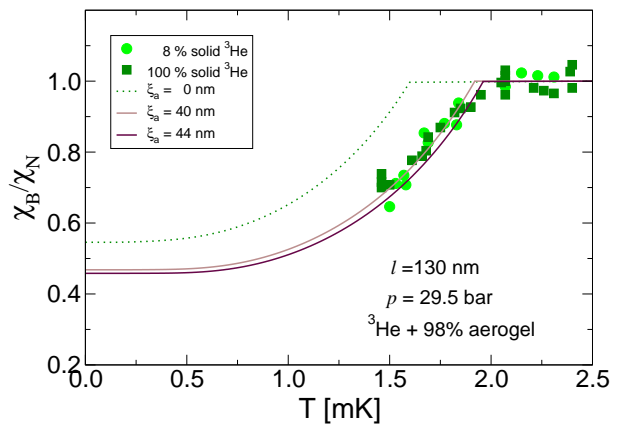}
\caption{\label{fig:GrenobleCompareChi29.5bar} Comparison to
experimental data for the magnetic susceptibility at
$p=29.5\,\text{bar}$. Only the liquid contribution to the
susceptibility is plotted for the two measurements with different
amounts of solid \He. The theoretical calculation is based on the
HSM and the dirty BW phase.}
\end{figure}

The magnetization of superfluid \He\ in $\approx 98\%$ aerogel was
measured by NMR in pure \He\ in aerogel and with varying amounts of
\Hefour\ pre-plating, and for intermediate ($p=17.0\,\text{bar}$)
and high  ($p=29.5\,\text{bar}$) pressures. We use the analysis of
the normal $^3$He susceptibility (shown in Fig. \ref{fig:Magnetiz})
to subtract the solid $^3$He component. In this pressure range the
effects of inhomogeneity in the aerogel are clearly reflected in
the weaker suppression of $T_c$ compared to the relative
suppression at lower pressures.\cite{sau03} Figures
\ref{fig:GrenobleCompareChi17bar} and
\ref{fig:GrenobleCompareChi29.5bar} show results for the magnetic
susceptibility of the ``dirty'' B-phase calculated for a mean free
path of $\ell = 130\,\text{nm}$. As is clearly shown $T_c$ as
predicted by the HSM model (i.e. $\xi_a=0$) is too strongly
suppressed. In addition the HSM predicts a larger increase in
the spin susceptibility for $T\to 0$, indicative of an overestimate
of the density of polarizable low-energy quasiparticles.

The effect of aerogel correlations, corresponding to typical void sizes of
$\xi_a\approx 40-44\,\text{nm}$, is shown in both figures. The
correlations lead to weaker suppression of $T_c$ compared to the HSM model
and to reduced pairbreaking. Both effects are clearly seen in the data for
both pressures, and are consistent with one another and with the mean free
path determined from the spin-diffusion data.


\end{document}